\input{aipcheck}

\documentclass[
    ,final            % use final for the camera ready runs
  ]
  {aipproc}

\layoutstyle{8x11single}
\def\Journal#1#2#3#4{{#1} {\bf #2}, #3 (#4)}

\def\RNC{\em Rivista Nuovo Cimento}

\def\NIMA{{\em Nucl. Instrum. Methods} A}

\def\PLB{{\em Phys. Lett.}  B}
\def\PRL{\em Phys. Rev. Lett.}
\def\PRD{{\em Phys. Rev.} D}

\def\GaC{\em Gravitation and Cosmology}
\def\GaCS{{\em Gravitation and Cosmology} Supplement}

\def\JETPL{\em JETP Lett.}
\def\PAN{\em Phys.Atom.Nucl.}
\def\CQG{\em Class. Quantum Grav.}
\def\APJ{\em Astrophys. J.}
\def\SCI{\em Science}
\def\MPLA{{\em Mod. Phys. Lett.}  A}
\def\IJTP{\em Int. J. Theor. Phys.}
\def\NJP{\em New J. of Phys.}
\def\JHEP{\em JHEP}
\def\BWP{\em Bled Workshops in Physics}
\def\s{{\,\rm s}}
\def\g{{\,\rm g}}
\def\eV{\,{\rm eV}}
\def\keV{\,{\rm keV}}
\def\MeV{\,{\rm MeV}}
\def\GeV{\,{\rm GeV}}
\def\TeV{\,{\rm TeV}}
\def\sv{\left<\sigma v\right>}
\def\({\left(}
\def\){\right)}
\def\cm{{\,\rm cm}}

\def\kpc{{\,\rm kpc}}
\def\beq{\begin{equation}}
\def\eeq{\end{equation}}
\def\bea{\begin{eqnarray}}
\def\eea{\end{eqnarray}}

\begin{document}

\title{The dark atoms of dark matter}


\author{Maxim Yu. Khlopov}{
  address={Centre for Cosmoparticle Physics "Cosmion" 115409 Moscow, Russia;\\
  National Research Nuclear University "Moscow Engineering Physics Institute", 115409 Moscow, Russia \\
    APC laboratory 10, rue Alice Domon et L\'eonie Duquet \\75205
Paris Cedex 13, France \\
khlopov@apc.univ-paris7.fr} }

\author{Andrey G. Mayorov} {
address={National Research Nuclear University (Moscow Engineering Physics Institute)\\
115409 Moscow, Russia\\
mayorov.a.g@gmail.com}}

\author{Evgeny Yu. Soldatov} {
address={National Research Nuclear University (Moscow Engineering Physics Institute)\\
115409 Moscow, Russia\\
Evgeny.Soldatov@cern.ch}}

\begin{abstract}
The nonbaryonic dark matter of the Universe is assumed to consist of
new stable particles.
 A specific case is possible, when new stable particles bear ordinary
 electric charge and bind in heavy "atoms" by ordinary Coulomb
 interaction. Such possibility is severely restricted by
 the constraints on anomalous isotopes of light elements that
 form positively charged heavy species with ordinary electrons.
 The trouble is avoided, if stable particles $X^{--}$ with charge
 -2 are in excess over their antiparticles (with charge +2) and there are no stable
 particles  with charges +1 and -1. Then primordial helium, formed in Big Bang
 Nucleosynthesis, captures all $X^{--}$ in
 neutral "atoms" of O-helium (OHe), thus creating a specific
Warmer than Cold nuclear-interacting composite dark matter. In the
Galaxy, destruction of OHe and acceleration of free $X^{--}$ can
result in anomalous component of cosmic rays. Collisions of OHe
atoms in the central part of Galaxy results in their excitation with
successive emission of electron-positron pairs, what can explain
excessive radiation of positron annihilation line, observed by
INTEGRAL. Slowed down in the terrestrial matter, OHe is elusive for
direct methods of underground dark matter detection based on the
search for effects of nuclear recoil in WIMP-nucleus collisions.
However OHe-nucleus interaction leads to their binding and in OHe-Na
system the energy of such level can be in the interval of energy 2-4
keV. The concentration of OHe in the matter of underground detectors
is rapidly adjusted to the incoming flux of cosmic O-helium.
Therefore the rate of energy release in radiative capture of Na by
OHe should experience annual modulations. It explains the results of
DAMA/NaI and DAMA/LIBRA experiments. The existence of low energy
bound state in OHe-Na system follows from the solution of
Schrodinger equation for relative motion of nucleus and OHe in a
spherically symmetrical potential, formed by the Yukawa tail of
nuclear scalar isoscalar attraction potential, acting on He beyond
the nucleus, and its Coulomb repulsion at distances from nuclear
surface, smaller than the size of OHe. Within the uncertainties of
nuclear physics parameters the values of coupling strength and mass
of sigma meson, mediating scalar isoscalar nuclear potential, are
found, at which the sodium nuclei have a few keV binding energy with
OHe. Transitions to more energetic levels of Na+OHe system imply
tunneling through Coulomb barrier that leads to suppression of
annual modulation of events with MeV-tens MeV energy release in the
correspondence with the results of DAMA experiments. The puzzles of
direct dark matter searches appear in this solution as a reflection
of nontrivial nuclear physics of OHe.
\end{abstract}

\maketitle

\section{Introduction}
 According to the modern cosmology, the dark matter, corresponding to
$25\%$ of the total cosmological density, is nonbaryonic and
consists of new stable particles. One can formulate the set of
conditions under which new particles can be considered as candidates
to dark matter (see e.g. \cite{book,Cosmoarcheology,Bled07} for
review and reference): they should be stable, saturate the measured
dark matter density and decouple from plasma and radiation at least
before the beginning of matter dominated stage. The easiest way to
satisfy these conditions is to involve neutral elementary weakly
interacting particles. However it is not the only particle physics
solution for the dark matter problem and more evolved models of
self-interacting dark matter are possible. In particular, new stable
particles may possess new U(1) gauge charges and bind by
Coulomb-like forces in composite dark matter species. Such dark
atoms would look nonluminous, since they radiate invisible light of
U(1) photons. In the studies of new particles Primordial Black holes can play the role of important theoretical tool (see \cite{pbh} for review and
references), which in particular can provide constraints on
particles with hidden gauge charges \cite{Dai:2009hx}.

 Here we consider composite dark matter scenarios, in which new
stable particles have ordinary electric charge, but escape
experimental discovery, because they are hidden in atom-like states
maintaining dark matter of the modern Universe. The main problem for
these scenarios is to suppress the abundance of positively charged
species bound with ordinary electrons, which behave as anomalous
isotopes of hydrogen or helium. This problem is unresolvable, if the
model predicts together with positively charged particles stable
particles $E^-$ with charge -1, as it is the case for tera-electrons
\cite{Glashow,Fargion:2005xz}. As soon as primordial helium is
formed in the Standard Big Bang Nucleosynthesis (SBBN) it captures
all the free $E^-$ in positively charged $(He E)^+$ ion, preventing
any further suppression of positively charged species. Therefore, in
order to avoid anomalous isotopes overproduction, stable particles
with charge -1 should be absent, so that stable negatively charged
particles should have charge -2 only.

Elementary particle frames for heavy stable -2 charged species are
provided by: (a) stable "antibaryons" $\bar U \bar U \bar U$ formed
by anti-$U$ quark of fourth generation \cite{Q,I,lom,Khlopov:2006dk}
(b) AC-leptons \cite{Khlopov:2006dk,5,FKS}, predicted in the
extension \cite{5} of standard model, based on the approach of
almost-commutative geometry \cite{bookAC}.  (c) Technileptons and
anti-technibaryons \cite{KK} in the framework of walking technicolor
models (WTC) \cite{Sannino:2004qp}. (d) Finally, stable charged
clusters $\bar u_5 \bar u_5 \bar u_5$ of (anti)quarks $\bar u_5$ of
5th family can follow from the approach, unifying spins and charges
\cite{Norma}. Since all these models also predict corresponding +2
charge antiparticles, cosmological scenario should provide mechanism
of their suppression, what can naturally take place in the
asymmetric case, corresponding to excess of -2 charge species,
$X^{--}$. Then their positively charged antiparticles can
effectively annihilate in the early Universe.

In all the models, in which new stable species belong to non-trivial
representations of electroweak SU(2) group sphaleron transitions at
high temperatures provide the relationship between baryon asymmetry
and excess of -2 charge stable species \cite{KK,KK2,unesco,iwara}.

 After it is formed
in the Standard Big Bang Nucleosynthesis (SBBN), $^4He$ screens the
$X^{--}$ charged particles in composite $(^4He^{++}X^{--})$ {\it
O-helium} ``atoms''
 \cite{I}.
 For different models of $X^{--}$ these "atoms" are also
called ANO-helium \cite{lom,Khlopov:2006dk}, Ole-helium
\cite{Khlopov:2006dk,FKS} or techni-O-helium \cite{KK}. We'll call
them all O-helium ($OHe$) in our further discussion, which follows
the guidelines of \cite{unesco,iwara,I2}.

In all these forms of O-helium, $X^{--}$ behaves either as lepton or
as specific "heavy quark cluster" with strongly suppressed hadronic
interaction. Therefore O-helium interaction with matter is
determined by nuclear interaction of $He$. These neutral primordial
nuclear interacting objects contribute to the modern dark matter
density and play the role of a nontrivial form of strongly
interacting dark matter \cite{Starkman,McGuire:2001qj,XQC}.

The active influence of this type of dark matter on nuclear
transformations needs special studies and development of OHe nuclear
physics. It is especially important for quantitative estimation of
role of OHe in Big Bang Nucleosynthesis and in stellar evolution.
This work is under way and its first results support the qualitative
picture of OHe cosmological evolution described in
\cite{I,FKS,KK,unesco,Khlopov:2008rp}.

Here after a brief review of main features of OHe Universe we
concentrate on its effects in underground detectors. We
qualitatively confirm the earlier guess
\cite{I,KK2,unesco,iwara,I2,Bled09} that the positive results of
dark matter searches in DAMA/NaI (see for review
\cite{Bernabei:2003za}) and DAMA/LIBRA \cite{Bernabei:2008yi}
experiments can be explained by effects of O-helium interaction with
the matter of underground detectors.

\section{Some features of O-helium Universe}

Following \cite{I,lom,Khlopov:2006dk,KK,unesco,iwara,I2} consider
charge asymmetric case, when excess of $X^{--}$ provides effective
suppression of positively charged species.

In the period $100\s \le t \le 300\s$  at $100 \keV\ge T \ge T_o=
I_{o}/27 \approx 60 \keV$, $^4He$ has already been formed in the
SBBN and virtually all free $X^{--}$ are trapped by $^4He$ in
O-helium ``atoms" $(^4He^{++} X^{--})$. Here the O-helium ionization
potential is\footnote{The account for charge distribution in $He$
nucleus leads to smaller value $I_o \approx 1.3 \MeV$
\cite{Pospelov}.} \beq I_{o} = Z_{x}^2 Z_{He}^2 \alpha^2 m_{He}/2
\approx 1.6 \MeV,\label{IO}\eeq where $\alpha$ is the fine structure
constant,$Z_{He}= 2$ and $Z_{x}= 2$ stands for the absolute value of
electric charge of $X^{--}$.  The size of these ``atoms" is
\cite{I,FKS} \beq R_{o} \sim 1/(Z_{x} Z_{He}\alpha m_{He}) \approx 2
\cdot 10^{-13} \cm \label{REHe} \eeq Here and further, if not
specified otherwise, we use the system of units $\hbar=c=k=1$.

O-helium, being an $\alpha$-particle with screened electric charge,
can catalyze nuclear transformations, which can influence primordial
light element abundance and cause primordial heavy element
formation. These effects need a special detailed and complicated
study. The arguments of \cite{I,FKS,KK,unesco,iwara} indicate that
this model does not lead to immediate contradictions with the
observational data. The conclusions that follow from our first steps
in the approach to OHe nuclear physics seem to support these
arguments.

Due to nuclear interactions of its helium constituent with nuclei in
the cosmic plasma, the O-helium gas is in thermal equilibrium with
plasma and radiation on the Radiation Dominance (RD) stage, while
the energy and momentum transfer from plasma is effective. The
radiation pressure acting on the plasma is then transferred to
density fluctuations of the O-helium gas and transforms them in
acoustic waves at scales up to the size of the horizon.

At temperature $T < T_{od} \approx 200 S^{2/3}_3\eV$ the energy and
momentum transfer from baryons to O-helium is not effective
\cite{I,KK} because $$n_B \sv (m_p/m_o) t < 1,$$ where $m_o$ is the
mass of the $OHe$ atom and $S_3= m_o/(1 \TeV)$. Here \beq \sigma
\approx \sigma_{o} \sim \pi R_{o}^2 \approx
10^{-25}\cm^2\label{sigOHe}, \eeq and $v = \sqrt{2T/m_p}$ is the
baryon thermal velocity. Then O-helium gas decouples from plasma. It
starts to dominate in the Universe after $t \sim 10^{12}\s$  at $T
\le T_{RM} \approx 1 \eV$ and O-helium ``atoms" play the main
dynamical role in the development of gravitational instability,
triggering the large scale structure formation. The composite nature
of O-helium determines the specifics of the corresponding dark
matter scenario.

At $T > T_{RM}$ the total mass of the $OHe$ gas with density $\rho_d
= (T_{RM}/T) \rho_{tot} $ is equal to
$$M=\frac{4 \pi}{3} \rho_d t^3 = \frac{4 \pi}{3} \frac{T_{RM}}{T} m_{Pl}
(\frac{m_{Pl}}{T})^2$$ within the cosmological horizon $l_h=t$. In
the period of decoupling $T = T_{od}$, this mass  depends strongly
on the O-helium mass $S_3$ and is given by \cite{KK}\beq M_{od} =
\frac{T_{RM}}{T_{od}} m_{Pl} (\frac{m_{Pl}}{T_{od}})^2 \approx 2
\cdot 10^{44} S^{-2}_3 \g = 10^{11} S^{-2}_3 M_{\odot}, \label{MEPm}
\eeq where $M_{\odot}$ is the solar mass. O-helium is formed only at
$T_{o}$ and its total mass within the cosmological horizon in the
period of its creation is $M_{o}=M_{od}(T_{od}/T_{o})^3 = 10^{37}
\g$.

On the RD stage before decoupling, the Jeans length $\lambda_J$ of
the $OHe$ gas was restricted from below by the propagation of sound
waves in plasma with a relativistic equation of state
$p=\epsilon/3$, being of the order of the cosmological horizon and
equal to $\lambda_J = l_h/\sqrt{3} = t/\sqrt{3}.$ After decoupling
at $T = T_{od}$, it falls down to $\lambda_J \sim v_o t,$ where $v_o
= \sqrt{2T_{od}/m_o}.$ Though after decoupling the Jeans mass in the
$OHe$ gas correspondingly falls down
$$M_J \sim v_o^3 M_{od}\sim 3 \cdot 10^{-14}M_{od},$$ one should
expect a strong suppression of fluctuations on scales $M<M_o$, as
well as adiabatic damping of sound waves in the RD plasma for scales
$M_o<M<M_{od}$. It can provide some suppression of small scale
structure in the considered model for all reasonable masses of
O-helium. The significance of this suppression and its effect on the
structure formation needs a special study in detailed numerical
simulations. In any case, it can not be as strong as the free
streaming suppression in ordinary Warm Dark Matter (WDM) scenarios,
but one can expect that qualitatively we deal with Warmer Than Cold
Dark Matter model.

Being decoupled from baryonic matter, the $OHe$ gas does not follow
the formation of baryonic astrophysical objects (stars, planets,
molecular clouds...) and forms dark matter halos of galaxies. It can
be easily seen that O-helium gas is collisionless for its number
density, saturating galactic dark matter. Taking the average density
of baryonic matter one can also find that the Galaxy as a whole is
transparent for O-helium in spite of its nuclear interaction. Only
individual baryonic objects like stars and planets are opaque for
it.

\section{Signatures of O-helium dark matter in the Galaxy}
The composite nature of O-helium dark matter results in a number of
observable effects, which we briefly discuss following
\cite{unesco}.
\subsection{Anomalous component of cosmic rays}
O-helium atoms can be destroyed in astrophysical processes, giving
rise to acceleration of free $X^{--}$ in the Galaxy.

O-helium can be ionized due to nuclear interaction with cosmic rays
\cite{I,I2}. Estimations \cite{I,Mayorov} show that for the number
density of cosmic rays $ n_{CR}=10^{-9}\cm^{-3}$ during the age of
Galaxy a fraction of about $10^{-6}$ of total amount of OHe is
disrupted irreversibly, since the inverse effect of recombination of
free $X^{--}$ is negligible. Near the Solar system it leads to
concentration of free $X^{--}$ $ n_{X}= 3 \cdot 10^{-10}S_3^{-1}
\cm^{-3}.$ After OHe destruction free $X^{--}$ have momentum of
order $p_{X} \cong \sqrt{2 \cdot M_{X} \cdot I_{o}} \cong 2 \GeV
S_3^{1/2}$ and velocity $v/c \cong 2 \cdot 10^{-3} S_3^{-1/2}$ and
due to effect of Solar modulation these particles initially can
hardly reach Earth \cite{KK2,Mayorov}. Their acceleration by Fermi
mechanism or by the collective acceleration forms power spectrum of
$X^{--}$ component at the level of $X/p \sim n_{X}/n_g = 3 \cdot
10^{-10}S_3^{-1},$ where $n_g \sim 1 \cm^{-3}$ is the density of
baryonic matter gas.

At the stage of red supergiant stars have the size $\sim 10^{15}
\cm$ and during the period of this stage$\sim 3 \cdot 10^{15} \s$,
up to $\sim 10^{-9}S_3^{-1}$ of O-helium atoms per nucleon can be
captured \cite{KK2,Mayorov}. In the Supernova explosion these OHe
atoms are disrupted in collisions with particles in the front of
shock wave and acceleration of free $X^{--}$ by regular mechanism
gives the corresponding fraction in cosmic rays. However, this
picture needs detailed analysis, based on the development of OHe
nuclear physics and numerical studies of OHe evolution in the
stellar matter.

If these mechanisms of $X^{--}$ acceleration are effective, the
anomalous low $Z/A$ component of $-2$ charged $X^{--}$ can be
present in cosmic rays at the level $X/p \sim n_{X}/n_g \sim
10^{-9}S_3^{-1},$ and be within the reach for PAMELA and AMS02
cosmic ray experiments.

In the framework of Walking Tachnicolor model the excess of both stable $X^{--}$
and $Y^{++}$ is possible \cite{KK2}, the latter being two-three orders of magnitude smaller, than the former.
It leads to the two-component composite dark matter scenario with the dominant OHe accompanied by a
subdominant WIMP-like component of $(X^{--}Y^{++})$ bound systems. Technibaryons and technileptons
can be metastable and decays of $X^{--}$ and $Y^{++}$ can provide explanation for anomalies,
observed in high energy cosmic positron spectrum by PAMELA and in high energy electron spectrum by FERMI and ATIC.

\subsection{Positron annihilation and gamma lines in galactic
bulge} Inelastic interaction of O-helium with the matter in the
interstellar space and its de-excitation can give rise to radiation
in the range from few keV to few  MeV. In the galactic bulge with
radius $r_b \sim 1 \kpc$ the number density of O-helium can reach
the value $n_o\approx 3 \cdot 10^{-3}/S_3 \cm^{-3}$ and the
collision rate of O-helium in this central region was estimated in
\cite{I2}: $dN/dt=n_o^2 \sigma v_h 4 \pi r_b^3 /3 \approx 3 \cdot
10^{42}S_3^{-2} \s^{-1}$. At the velocity of $v_h \sim 3 \cdot 10^7
\cm/\s$ energy transfer in such collisions is $\Delta E \sim 1 \MeV
S_3$. These collisions can lead to excitation of O-helium. If 2S
level is excited, pair production dominates over two-photon channel
in the de-excitation by $E0$ transition and positron production with
the rate $3 \cdot 10^{42}S_3^{-2} \s^{-1}$ is not accompanied by
strong gamma signal. According to \cite{Finkbeiner:2007kk} this rate
of positron production for $S_3 \sim 1$ is sufficient to explain the
excess in positron annihilation line from bulge, measured by
INTEGRAL (see \cite{integral} for review and references). If $OHe$
levels with nonzero orbital momentum are excited, gamma lines should
be observed from transitions ($ n>m$) $E_{nm}= 1.598 \MeV (1/m^2
-1/n^2)$ (or from the similar transitions corresponding to the case
$I_o = 1.287 \MeV $) at the level $3 \cdot 10^{-4}S_3^{-2}(\cm^2 \s
\MeV ster)^{-1}$.

\section{O-helium in the terrestrial matter}
The evident consequence of the O-helium dark matter is its
inevitable presence in the terrestrial matter, which appears opaque
to O-helium and stores all its in-falling flux.

Such neutral $(^4He^{++}X^{--})$ ``atoms" may provide a catalysis of
cold nuclear reactions in ordinary matter (much more effectively
than muon catalysis). This effect needs a special and thorough
investigation. On the other hand, $X^{--}$ capture by nuclei,
heavier than helium, can lead to production of anomalous isotopes,
but the arguments, presented in \cite{I,FKS,KK} indicate that their
abundance should be below the experimental upper limits.

It should be noted that the nuclear cross section of the O-helium
interaction with matter escapes the severe constraints
\cite{McGuire:2001qj} on strongly interacting dark matter particles
(SIMPs) \cite{Starkman,McGuire:2001qj} imposed by the XQC experiment
\cite{XQC}. Therefore, a special strategy of direct O-helium  search
is needed, as it was proposed in \cite{Belotsky:2006fa}.

After they fall down terrestrial surface the in-falling $OHe$
particles are effectively slowed down due to elastic collisions with
matter. Then they drift, sinking down towards the center of the
Earth with velocity \beq V = \frac{g}{n \sigma v} \approx 80 S_3
A^{1/2} \cm/\s. \label{dif}\eeq Here $A \sim 30$ is the average
atomic weight in terrestrial surface matter, $n=2.4 \cdot 10^{24}/A$
is the number of terrestrial atomic nuclei, $\sigma v$ is the rate
of nuclear collisions and $g=980~ \cm/\s^2$.

Near the Earth's surface, the O-helium abundance is determined by
the equilibrium between the in-falling and down-drifting fluxes.

The in-falling O-helium flux from dark matter halo is
$$
  F=\frac{n_{0}}{8\pi}\cdot |\overline{V_{h}}+\overline{V_{E}}|,
$$
where $V_{h}$-speed of Solar System (220 km/s), $V_{E}$-speed of
Earth (29.5 km/s) and $n_{0}=3 \cdot 10^{-4} S_3^{-1} \cm^{-3}$ is the
local density of O-helium dark matter. Here, for qualitative estimation,
we don't take into account velocity dispersion and distribution of particles
in the incoming flux that can lead to significant effect.

At a depth $L$ below the Earth's surface, the drift timescale is
$t_{dr} \sim L/V$, where $V \sim 400 S_3 \cm/\s$ is given by
Eq.~(\ref{dif}). It means that the change of the incoming flux,
caused by the motion of the Earth along its orbit, should lead at
the depth $L \sim 10^5 \cm$ to the corresponding change in the
equilibrium underground concentration of $OHe$ on the timescale
$t_{dr} \approx 2.5 \cdot 10^2 S_3^{-1}\s$.

In underground detectors, $OHe$ ``atoms'' are slowed down to thermal
energies and give rise to energy transfer $\sim 2.5 \cdot 10^{-4}
\eV A/S_3$, far below the threshold for direct dark matter
detection. It makes this form of dark matter insensitive to the
severe CDMS \cite{Akerib:2005kh} and XENON100 constraints
\cite{xenon}. However, $OHe$ reactions with the matter of
underground detectors can lead to observable effects.

The equilibrium concentration, which is established in the matter of
underground detectors, is given by
\begin{equation}
    n_{oE}=\frac{2\pi \cdot F}{V} = n_{oE}^{(1)}+n_{oE}^{(2)}\cdot sin(\omega (t-t_0)),
    \label{noE}
\end{equation}
where $\omega = 2\pi/T$, $T=1yr$ and $t_0$ is the phase. The
averaged concentration is given by
\begin{equation}
    n_{oE}^{(1)}=\frac{n_o}{320S_3 A_{med}^{1/2}} V_{h}
\end{equation}
and the annual modulation of concentration is characterized by
\begin{equation}
    n_{oE}^{(2)}= \frac{n_o}{640S_3 A_{med}^{1/2}} V_E
\end{equation}

The rate of nuclear reactions of OHe with nuclei is proportional to
the local concentration and the energy release in these reactions
should lead to observable signal. There are two parts of the signal:
the one determined by the constant part and annual modulation, which
is concerned by the strategy of dark matter search in DAMA
experiment \cite{Bernabei:2008yi}.

%%%%%%%%%%%%%%%%%%%%%%%%%%%%%%%%%%%%%%%%%%%%%%%%%%%%%%%%%%%%%%%%%%%%%%%%

\section{Low energy bound state of O-helium with nuclei}

Our explanation \cite{unesco,iwara,Bled09} of the results of
DAMA/NaI or DAMA/LIBRA experiments is based on the idea that OHe,
slowed down in the matter of detector, can form a few keV bound
state with nucleus, in which OHe is situated \textbf{beyond} the
nucleus. Therefore the positive result of these experiments is
explained by reaction
\begin{equation}
A+(^4He^{++}X^{--}) \rightarrow [A(^4He^{++}X^{--})]+\gamma
\label{HeEAZ}
\end{equation}
with nuclei in DAMA detector. In our earlier studies
\cite{unesco,iwara,Bled09} the conditions were found, under which
both sodium and iodine nuclei have a few keV bound states with OHe,
explaining the results of DAMA experiments by OHe radiative capture
to these levels. Here we extend the set of our solutions by the
case, when the results of DAMA experiment can be explained by
radiative OHe capture by sodium only and there are no such bound
states with iodine and Tl.

\subsection{Low energy bound state of O-helium with nuclei}

Schroedinger equation for OHe-nucleus system is reduced (taking apart
the equation for the center of mass) to the equation of relative
motion for the reduced mass
\begin{equation}
            m=\frac{Am_p m_o}{Am_p+m_o},
            \label{m}
 \end{equation}
where $m_p$ is the mass of proton and $m_o\approx M_X+4m_p$ is the
mass of OHe. Since $m_o \approx M_X \gg A m_p$, center of mass of
Ohe-nucleus system approximately coincides with the position of
$X^{--}$.

In the case of orbital momentum \emph{l}=0 the wave functions depend
only on \emph{r}.

The approach of \cite{unesco,iwara,Bled09,Bled08,Levels,DMDA} assumes the following
picture: at the distances larger, than its size, OHe is neutral,
being only the source of a Coulomb field of $X^{--}$ screened by
$He$ shell
\begin{equation}
U_c= \frac{Z_{X} Z \alpha  \cdot F_X(r)}{r}, \label{epotem}
\end{equation}
where $Z_{X}=-2$ is the charge of $X^{--}$, $Z$ is charge of
nucleus, $F_X(r)=(1+r/r_o) exp(-2r/r_o)$ is the screening factor of
Coulomb potential (see e.g.\cite{LL3}) of $X^{--}$ and $r_o$ is the
size of OHe. Owing to the negative sign of $Z_{X}=-2$, this
potential provides attraction of nucleus to OHe.

Then helium shell of OHe starts to feel Yukawa exponential tail of
attraction of nucleus to $He$ due to scalar-isoscalar nuclear
potential. It should be noted that scalar-isoscalar nature of He
nucleus excludes its nuclear interaction due to $\pi$ or $\rho$
meson exchange, so that the main role in its nuclear interaction
outside the nucleus plays $\sigma$ meson exchange, on which nuclear
physics data are not very definite. The nuclear potential depends on
the relative distance between He and nucleus and we take it in the
form
\begin{equation}
U_n=-\frac{A_{He} A g^2 exp{(-\mu
|\vec{r}-\vec{\rho}|)}}{|\vec{r}-\vec{\rho}|}. \label{epotnuc}
\end{equation}
Here $\vec{r}$ is radius vector to nucleus, $\vec{\rho}$ is the
radius vector to He in OHe, $A_{He}=4$ is atomic weight of helium,
$A$ is atomic weight of nucleus, $\mu$ and $g^2$ are the mass and
coupling of $\sigma$ meson - mediator of nuclear attraction.

Strictly speaking, starting from this point we should deal with a
three-body problem for the system of He, nucleus and $X^{--}$ and
the correct quantum mechanical description should be based on the
cylindrical and not spherical symmetry. In the present work we use
the approximation of spherical symmetry and take into account
nuclear attraction beyond the nucleus in a two different ways: 1)
nuclear attraction doesn't influence the structure of OHe, so that
the Yukawa potential (\ref{epotnuc}) is averaged over $|\vec{\rho}|$
for spherically symmetric wave function of He shell in OHe; 2)
nuclear attraction changes the structure of OHe so that He takes the
position $|\vec{\rho}|=r_o$, which is most close to the nucleus. Due
to strong attraction of He by the nucleus the second case (which is
labeled "b" in successive numerical calculations) seems more
probable. In the lack of the exact solution of the problem we
present both the results, corresponding to the first case (which are
labeled "m" in successive numerical calculations), and to the second
case (which is labeled "b") in order to demonstrate high sensitivity
of the numerical results to choice of parameters.

In the both cases nuclear attraction results in the polarization of
OHe and the mutual attraction of nucleus and OHe is changed by
Coulomb repulsion of $He$ shell. Taking into account Coulomb
attraction of nucleus by $X^{--}$ one obtains dipole Coulomb barrier
of the form
\begin{equation}
U_d=\frac{Z_{He} Z \alpha r_o}{r^2}. \label{epotdip}
\end{equation}

When helium is completely merged with the nucleus the interaction is
reduced to the oscillatory potential (\ref{potosc}) of $X^{--}$ with
homogeneously charged merged nucleus with the charge $Z+2$, given by
\begin{equation}
    E_m=\frac{3}{2}(\frac{(Z+2) Z_x \alpha}{R}-\frac{1}{R}(\frac{(Z+2) Z_x \alpha}{(A+4) m_p R})^{1/2}).
\label{potosc}
\end{equation}

\begin{figure}
            \includegraphics[width=4in]{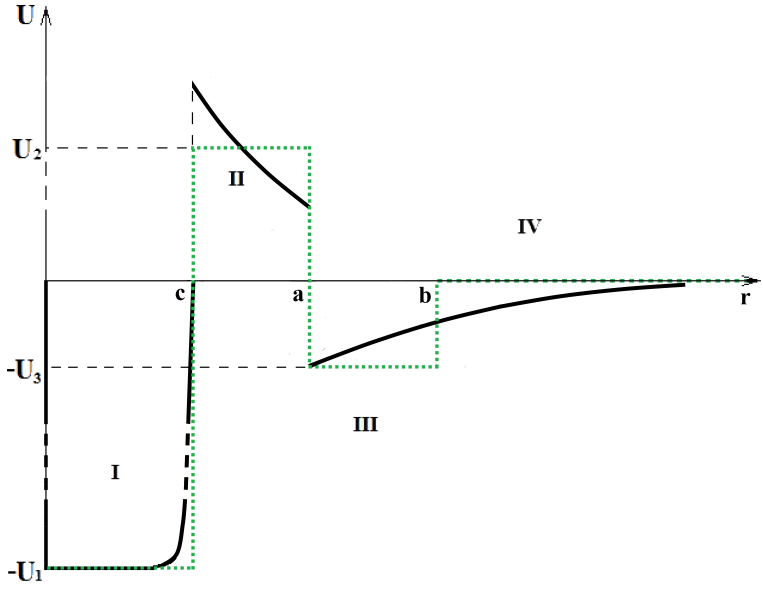}\\
        \caption{The approximation of rectangular well for potential of OHe-nucleus system.}\label{pic1}
    \end{figure}

To simplify the solution of Schroedinger equation we approximate the
potentials (\ref{epotem})-(\ref{potosc}) by a rectangular potential
that consists of a potential well with the depth $U_1$ at $r<c=R$,
where $R$ is the radius of nucleus, of a rectangular dipole Coulomb
potential barrier $U_2$ at $R \le r<a=R+r_o+r_{he}$, where $r_{he}$
is radius of helium nucleus, and of the outer potential well $U_3$,
formed by the Yukawa nuclear interaction (\ref{epotnuc}) and
residual Coulomb interaction (\ref{epotem}). The values of $U_1$ and
$U_2$ were obtained by the averaging of the (\ref{potosc}) and
(\ref{epotdip}) in the corresponding regions, while $U_3$ was equal
to the value of the nuclear potential (\ref{epotnuc}) at $r=a$ and
the width of this outer rectangular well (position of the point b)
was obtained by the integral of the sum of potentials
(\ref{epotnuc}) and (\ref{epotem}) from $a$ to $\infty$.
 It leads to the approximate potential, presented on Fig. \ref{pic1}.

Solutions of Schroedinger equation for each of the four regions,
indicated on Fig. \ref{pic1}, are given in textbooks (see
e.g.\cite{LL3}) and their sewing determines the condition, under
which a low-energy  OHe-nucleus bound state appears in the region
III.

\begin{figure}
            \includegraphics[width=6in]{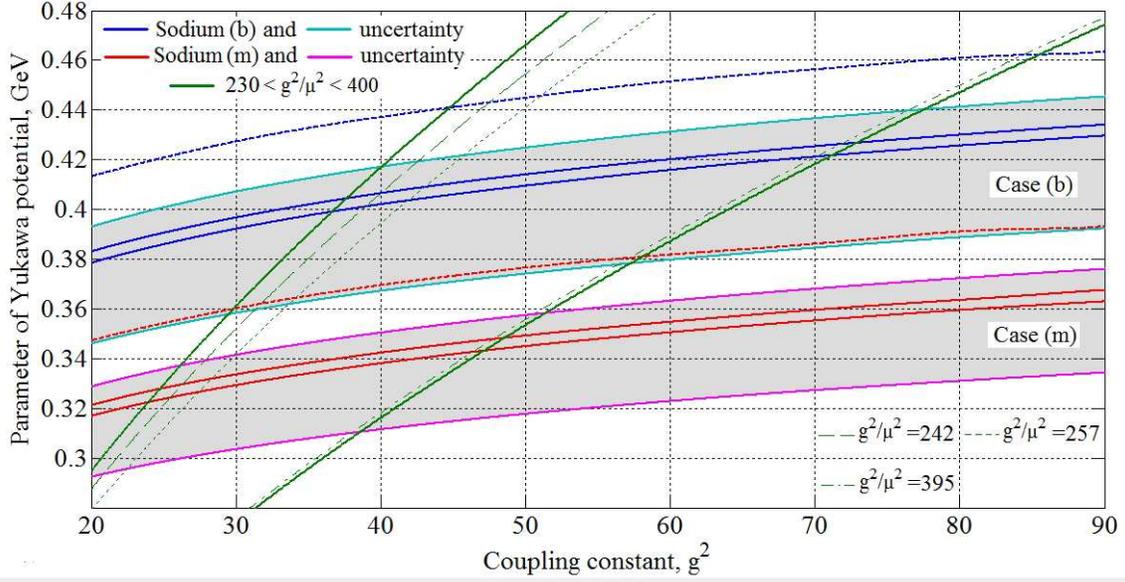}\\
        \caption{
        The region of parameters $\mu$ and $g^2$, for which Na-OHe system has a level in the interval 4 keV.
        Two lines determine at $d_o=1.2/(200 \MeV)$ the region of parameters, at which the bound
        system of this element with OHe has a 4 keV level.
        In the region between the two strips the energy of level is below 4 keV.
        There are also indicated the range of $g^2/\mu^2$ (dashed lines) as well as their preferred values
        (thin lines) determined in \cite{nuclear} from parametrization of the relativistic ($\sigma-\omega$) model for nuclear matter.
        The uncertainty in the determination of parameter $1.15/(200 \MeV)<d_o <1.3/(200 \MeV)$ results in the
        uncertainty of $\mu$ and $g^2$ shown by the shaded regions surrounding the lines.
        The case of nuclear Yukawa potential $U_{3m}$, averaged over the orbit of He in OHe, corresponds to the
        lower lines and shaded region, while the upper lines and shaded region around them illustrate
        the case of nuclear Yukawa potential $U_{3b}$ with the position
        of He most close to the nucleus at $\rho=r_o$.}\label{Na}
   \end{figure}

The energy of this bound state and its existence strongly depend on
the parameters $\mu$ and $g^2$ of nuclear potential (\ref{epotnuc}).
On the Fig. \ref{Na} the regions of these parameters, giving 4 keV
energy level in OHe bound state with sodium are presented. Radiative
capture to this level can explain results of DAMA/NaI and DAMA/LIBRA
experiments with the account for their energy resolution
\cite{DAMAlibra}. The lower shaded region on Fig. \ref{Na}
corresponds to the case of nuclear Yukawa potential $U_{3m}$,
averaged over the orbit of He in OHe, while the upper region
corresponds to the case of nuclear Yukawa potential $U_{3b}$ with
the position of He most close to the nucleus at $\rho=r_o$.The
result is also sensitive to the precise value of $d_o$, which
determines the size of nuclei $R=d_o A^{1/3}$. The two narrow strips
in each region correspond to the experimentally most probable value
$d_o=1.2/(200 \MeV)$. In these calculations the mass of OHe was
taken equal to $m_o=1 TeV$, however the results weakly depend on the
value of $m_o>1 TeV$.

It is interesting that the values of $\mu$ on Fig. \ref{Na} are compatible with the results of recent experimental measurements of mass of sigma meson ($\mu \sim 404 \MeV$) \cite{sigma}.

The rate of radiative capture of OHe by nuclei can be calculated
\cite{unesco,iwara} with the use of the analogy with the radiative
capture of neutron by proton with the account for: i) absence of M1
transition that follows from conservation of orbital momentum and
ii) suppression of E1 transition in the case of OHe. Since OHe is
isoscalar, isovector E1 transition can take place in OHe-nucleus
system only due to effect of isospin nonconservation, which can be
measured by the factor $f = (m_n-m_p)/m_N \approx 1.4 \cdot
10^{-3}$, corresponding to the difference of mass of neutron,$m_n$,
and proton,$m_p$, relative to the mass of nucleon, $m_N$. In the
result the rate of OHe radiative capture by nucleus with atomic
number $A$ and charge $Z$ to the energy level $E$ in the medium with
temperature $T$ is given by
\begin{equation}
    \sigma v=\frac{f \pi \alpha}{m_p^2} \frac{3}{\sqrt{2}} (\frac{Z}{A})^2 \frac{T}{\sqrt{Am_pE}}.
    \label{radcap}
\end{equation}

Formation of OHe-nucleus bound system leads to energy release of its
binding energy, detected as ionization signal.  In the context of
our approach the existence of annual modulations of this signal in
the range 2-6 keV and absence of such effect at energies above 6 keV
means that binding energy of Na-OHe system in DAMA experiment should
not exceed 6 keV, being in the range 2-4 keV. The amplitude of
annual modulation of ionization signal (measured in counts per day
per kg, cpd/kg) is given by
\begin{equation}
\zeta=\frac{3\pi \alpha \cdot n_o N_A V_E t Q}{640\sqrt{2}
A_{med}^{1/2} (A_I+A_{Na})} \frac{f}{S_3 m_p^2} (\frac{Z_i}{A_i})^2
\frac{T}{\sqrt{A_i m_p E_i}}= a_i\frac{f}{S_3^2} (\frac{Z_i}{A_i})^2
\frac{T}{\sqrt{A_i m_p E_i}}. \label{counts}
\end{equation}
Here $N_A$ is Avogadro number, $i$ denotes Na, for which numerical
factor $a_i=4.3\cdot10^{10}$, $Q=10^3$ (corresponding to 1kg of the
matter of detector), $t=86400 \s$, $E_i$ is the binding energy of
Na-OHe system and $n_{0}=3 \cdot 10^{-4} S_3^{-1} \cm^{-3}$ is the
local density of O-helium dark matter near the Earth. The value of
$\zeta$ should be compared with the integrated over energy bins
signals in DAMA/NaI and DAMA/LIBRA experiments and the result of
these experiments can be reproduced for $E_{Na} = 3 \keV$. The
account for energy resolution in DAMA experiments \cite{DAMAlibra}
can explain the observed energy distribution of the signal from
monochromatic photon (with $E_{Na} = 3 \keV$) emitted in OHe
radiative capture.

At the corresponding values of $\mu$ and $g^2$ there is no binding
of OHe with iodine and thallium.

It should be noted that the results of DAMA experiment exhibit also absence
of annual modulations at the energy of MeV-tens MeV. Energy release in this range should take place,
if OHe-nucleus system comes to the deep level inside the nucleus (in the region I of Fig. \ref{pic1}).
This transition implies tunneling through dipole Coulomb barrier and is suppressed below the experimental limits.
%Preliminary results give the energy level of
%for $\mu= 320 \MeV$ and $g^2=2$, $\mu= 350 \MeV$ and $g^2=4$, $\mu= 380 \MeV$ and $g^2=10$ or for $\mu= 460 \MeV$ and $g^2=100$.

\subsection{OHe radiative capture by other nuclei}

The important qualitative feature of the presented solution is the
restricted range of intermediate nuclei, in which the OHe-nucleus
state beyond nuclei is possible. For the chosen range of nuclear
parameters, reproducing the results of DAMA/NaI and DAMA/LIBRA, we
can calculate the binding energy of OHe-nucleus states in nuclei,
corresponding to chemical composition of set-ups in other
experiments. It turns out that  there are no such states for light
and heavy nuclei. In the case of nuclear Yukawa potential $U_{3b}$,
corresponding to the position of He most close to the nucleus at
$\rho=r_o$, the range of nuclei with bound states with OHe
corresponds to the part of periodic table between B and Ti. This
result is stable independent on the used scheme of numerical
calculations. In the case of potential $U_{3m}$,
averaged over the orbit of He in OHe, there are no OHe bound states
with nuclei, lighter than Be and heavier than Ti. However, the
results are very sensitive to the numerical factors of calculations
and the existence of OHe-Ge and OHe-Ga bound states at a narrow
window of parameters $\mu$ and $g^2$ turns to be strongly dependent
on these factors so that change in numbers smaller than 1\% can give
qualitatively  different result for Ge and Ga.
Both for the cases (b) and (m) there is a stable conclusion that
there are no OHe-nucleus bound states with Xe, I and Tl.

\begin{figure}
            \includegraphics[width=6in]{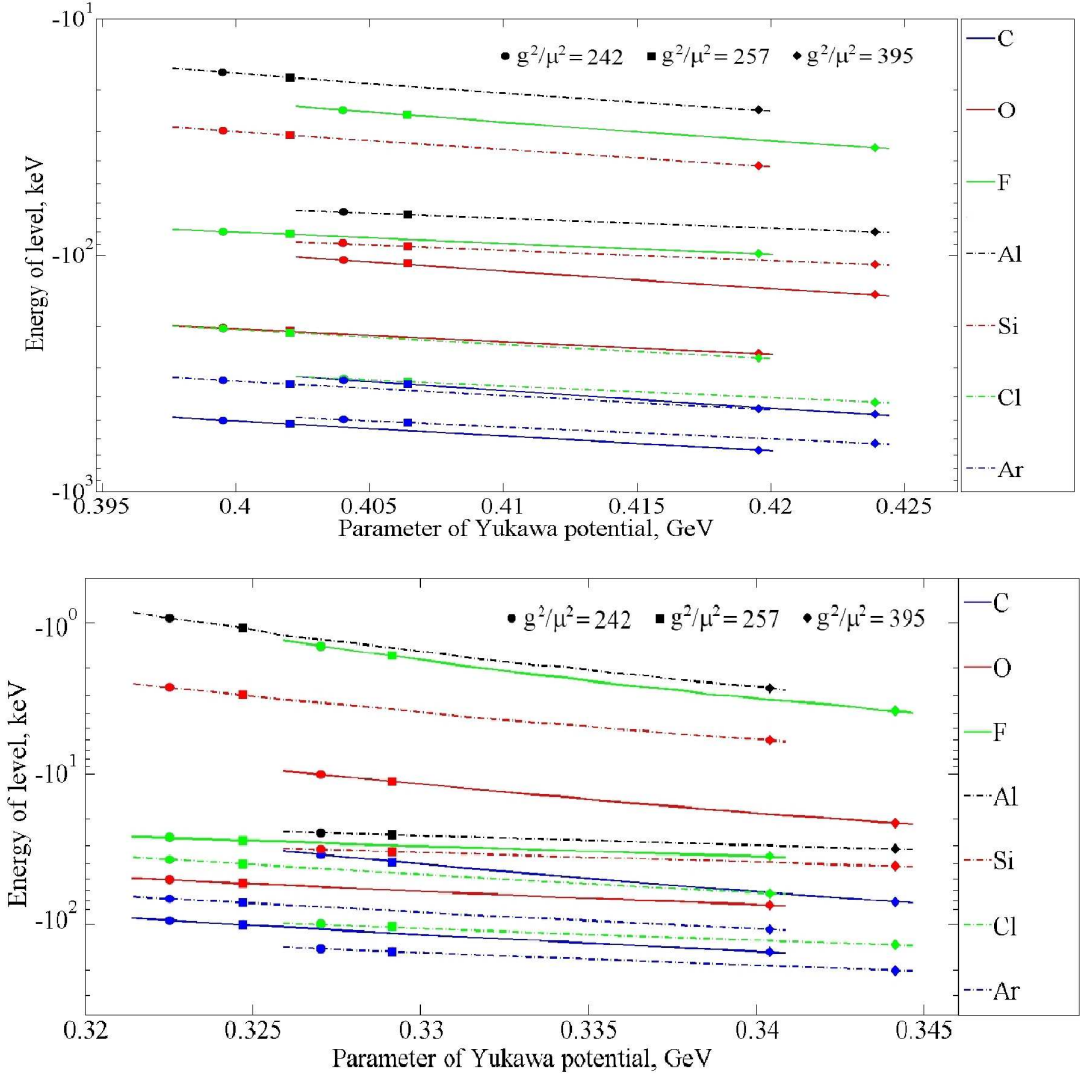}\\
        \caption{Energy levels in OHe bound system with carbon,
        oxygen, fluorine, argon, silicon, aluminium and chlorine for the case of the nuclear
Yukawa potential $U_{3b}$ (upper figure) and $U_{3m}$ (lower figure). The predictions are given for the range
of $g^2/\mu^2$ determined in \cite{nuclear} from parametrization of
the relativistic ($\sigma-\omega$) model for nuclear matter. The
preferred values of $g^2/\mu^2$ are indicated by the corresponding
marks (squares or circles).}\label{elements}
   \end{figure}

For the experimentally preferred value $d_o=1.2/(200 \MeV)$ the
results of calculation of the binding energy of OHe-nucleus systems
for carbon, oxygen, fluorine, argon, silicon, aluminium and chlorine
are presented on Fig. \ref{elements} for the case of the nuclear
Yukawa potential $U_{3b}$ (upper figure) and $U_{3m}$ (lower figure). The difference in these results
demonstrates their high sensitivity to the choice of parameters.
For the parameters, reproducing results of DAMA experiment the
predicted energy level of OHe-silicon bound state is generally
beyond the range 2-6 keV, being in the most cases in the range of
30-40 keV or 90-110 keV by absolute value. It makes elusive a
possibility to test DAMA results by search for ionization signal in
the same range 2-6 keV in other set-ups with content that differs
from Na and I. Even in the extreme case (m) of ionization signal in
the range 2-6 keV our approach naturally predicts its suppression in
accordance with the results of CDMS \cite{Kamaev:2009gp}.

It should be noted that strong sensitivity of the existence of the
OHe-Ge bound state to the values of numerical factors \cite{Levels}
doesn't exclude such state for some window of nuclear physics
parameters. The corresponding binding energy  would be about 450-460
keV, what proves the above statement even in that case.

Since OHe capture rate is proportional to the temperature, it looks
like it is suppressed in cryogenic detectors by a factor of order
$10^{-4}$. However, for the size of cryogenic devices  less, than
few tens meters, OHe gas in them has the thermal velocity of the
surrounding matter and the suppression relative to room temperature
is only $\sim m_A/m_o$. Then the rate of OHe radiative capture in
cryogenic detectors is given by Eq.(\ref{radcap}), in which room
temperature $T$ is multiplied by factor $m_A/m_o$, and the
ionization signal (measured in counts per day per kg, cpd/kg) is
given by Eq.(\ref{counts}) with the same correction for $T$
supplemented by additional factors $2 V_h/V_E$ and
$(A_I+A_{Na})/A_i$, where $i$ denotes Si.
 To illustrate possible effects of OHe in
various cryogenic detectors we give in Tables~\ref{ta1} and
\ref{ta2} energy release, radiative capture rate and counts per day
per kg for the pure silicon for the preferred values of nuclear
parameters.

\begin{table}
\caption{Effects of OHe in pure silicon cryogenic detector in the
case m for nuclear Yukawa potential $U_{3m}$, averaged over the
orbit of He in OHe \cite{Levels}.}

\begin{tabular}{|c|c|c|c|c|c|c|}
    \hline
        $g^2/\mu^2, GeV^{-1}$ & 242 & 242 & 257 & 257 & 395 & 395\\
    \hline
        Energy, $keV$ & 2.7 & 31.9 & 3.0 & 33.2 & 6.1 & 41.9\\
    \hline
        $\sigma V \cdot 10^{-33}, cm^3/s$ & 19.3 & 5.6 & 18.3 & 5.5 & 12.8 & 4.9\\
    \hline
        $\xi \cdot 10^{-2}, cpd/kg$ & 10.8 & 3.1 & 10.2 & 3.1 & 7.2 & 2.7\\
    \hline
\end{tabular}\label{ta1}

\end{table}

\begin{table}
\caption{Effects of OHe in pure silicon cryogenic detector for the
case of the nuclear Yukawa potential $U_{3b}$ with the position of
He most close to the nucleus \cite{Levels}.}

\begin{tabular}{|c|c|c|c|c|c|c|}
    \hline
        $g^2/\mu^2, GeV^{-1}$ & 242 & 242 & 257 & 257 & 395 & 395\\
    \hline
        Energy, $keV$ & 29.8 & 89.7 & 31.2 & 92.0 & 42.0 & 110.0\\
    \hline
        $\sigma V \cdot 10^{-33}, cm^3/s$ & 5.8 & 3.3 & 5.7 & 3.3 & 4.9 & 3.0\\
    \hline
        $\xi \cdot 10^{-2}, cpd/kg$ & 3.3 & 1.9 & 3.2 & 1.9 & 2.7 & 1.7\\
    \hline
\end{tabular}\label{ta2}

\end{table}

\section{Conclusions}
To be present in the modern Universe, dark matter should survive to
the present time. It assumes a conservation law for dark matter
particles, what implies these particles to possess some new
fundamental symmetry and corresponding conserved charge. If
particles possess new gauge U(1) symmetry, they can bind by the
corresponding Coulomb-like interaction in dark atoms, emitting dark
U(1) photons. Here, we have studied a possibility that dark matter
particles possess ordinary electric charge and are bound in dark
atoms by ordinary Coulomb interaction. However restricted, this
possibility is not excluded by observations. The existence of heavy
stable charged particles may not only be compatible with the
experimental constraints but can even lead to composite dark matter
scenario of nuclear interacting Warmer than Cold Dark Matter. This
new form of dark matter can provide explanation of excess of
positron annihilation line radiation, observed by INTEGRAL in the
galactic bulge.

In the first three minutes primordial stable particles with charge -2 are bound with helium so that their specific features are shielded by a nuclear interacting helium shell that determines their successive evolution in the form of dark OHe atoms. Detailed analysis of this evolution implies development of OHe nuclear physics that is now under way. Its first results indicate that OHe experiences dominantly elastic collisions with the matter and that its inelastic collisions with capture inside nuclei are strongly suppressed. It provides qualitative proof of earlier guess that formation of anomalous isotopes due to OHe capture is within the experimental upper limits.

Destruction of OHe, e.g. by cosmic rays or in SN explosions gives rise to free $X^{--}$, but relative amount of such free X as compared with those bound in OHe is at least by factor $10^7$ smaller in the Galaxy and their penetration in the Solar System is hindered by the Solar wind. We qualitatively expect existence of an X component of cosmic rays, but of course we need more studies to check other possible signatures of free X. The search for stable -2 charge component of cosmic
rays is challenging for PAMELA and AMS02 experiments. Decays of
heavy charged constituents of composite dark matter can provide
explanation for anomalies in spectra of cosmic high energy positrons
and electrons, observed by PAMELA, FERMI and ATIC.

In the surrounding terrestrial matter OHe concentration is regulated by the equilibrium between the down streaming diffusion to the center of the Earth and the incoming cosmic flux. As it follows from Eq. (7) this averaged equilibrium concentration is of the order of few species per $\cm^3$. Effects of these species in the matter need, certainly, special studies but in the first approximation it seems that OHe signatures in the underground experiments, discussed in the present paper,  provide their most sensitive probe.

In the context of
our approach search for heavy stable charged quarks and leptons at
LHC acquires the significance of experimental probe for components
of cosmological composite dark matter. Accelerator test for composite $X^{--}$ can be only indirect:
in the lack of possibility of direct search for triple heavy quark or
antiquark states it is possible to search for heavy hadrons, composed of single $U$ or
$\bar U$ and light quarks. One can analyze mass spectrum of these
hadrons and find the correlations, by which they can be in principle
detected and discriminated from other similar particles, like R-hadrons.

The way to search for AC leptons and techniparticles looks much more
straightforward. The set of experimental signatures for these particles
can provide their clear distinction from other hypothetical exotic particles.

The results of dark matter search in experiments DAMA/NaI and
DAMA/LIBRA can be explained in the framework of our scenario without
contradiction with the results of other groups. This scenario can be
realized in different frameworks, in particular, in the extensions
of Standard Model, based on the approach of almost commutative
geometry, in the model of stable quarks of 4th generation that can
be naturally embedded in the heterotic superstring phenomenology, in
the models of stable technileptons and/or techniquarks, following
from Minimal Walking Technicolor model or in the approach unifying
spin and charges. Our approach contains distinct features, by which
the present explanation can be distinguished from other recent
approaches to this problem \cite{Edward} (see also for review and
more references in \cite{Gelmini}).

The proposed explanation is based on the mechanism of low energy
binding of OHe with nuclei. Within the uncertainty of nuclear
physics parameters there exists a range at which OHe binding energy
with sodium is in the interval 2-4 keV. Radiative capture of OHe to
this bound state leads to the corresponding energy release observed
as an ionization signal in DAMA detector.

OHe concentration in the matter of underground detectors is
determined by the equilibrium between the incoming cosmic flux of
OHe and diffusion towards the center of Earth. It is rapidly
adjusted and follows the change in this flux with the relaxation
time of few minutes. Therefore the rate of radiative capture of OHe
should experience annual modulations reflected in annual modulations
of the ionization signal from these reactions.
%The method to calculate the rate of OHe reactions was developed and
%the calculated total amount of such events is shown to be consistent
%with the results of DAMA/NaI and DAMA/LIBRA experiments for the mass
%of OHe around 1 TeV. This method can be applied to the analysis of
%the whole set of inelastic processes, induced by O-helium in matter.

An inevitable consequence of the proposed explanation is appearance
in the matter of DAMA/NaI or DAMA/LIBRA detector anomalous
superheavy isotopes of sodium, having the mass roughly by $m_o$
larger, than ordinary isotopes of these elements. If the atoms of
these anomalous isotopes are not completely ionized, their mobility
is determined by atomic cross sections and becomes about 9 orders of
magnitude smaller, than for O-helium. It provides their conservation
in the matter of detector. Therefore mass-spectroscopic analysis of
this matter can provide additional test for the O-helium nature of
DAMA signal. Methods of such analysis should take into account the
fragile nature of OHe-Na bound states, since their binding energy is
only few keV.

With the account for high sensitivity of the numerical results to
the values of nuclear parameters and for the approximations, made in
the calculations, the presented results can be considered only as an
illustration of the possibility to explain puzzles of dark matter
search in the framework of composite dark matter scenario. An
interesting feature of this explanation is a conclusion that the
ionization signal expected in detectors with the content, different
from NaI, can be dominantly in the energy range beyond 2-6 keV.
Moreover, it can be absent in detectors, containing heavy nuclei
(e.g. xenon). Therefore test of results of DAMA/NaI and DAMA/LIBRA
experiments by other experimental groups can become a very
nontrivial task.

Our results show that the ionization signal, detected by DAMA, may
be absent in detectors containing light elements. In particular,
there is predicted no low-energy binding of OHe with $^3He$ and
correspondingly no ionization signal in keV range in the designed
$^3He$ dark matter detectors. Therefore development of experimental
methods of dark matter detection will extend the possibilities to
test hypothesis of composite dark matter.

It is interesting to note that in the framework of our approach
positive result of experimental search for WIMPs by effect of their
nuclear recoil would be a signature for a multicomponent nature of
dark matter. Such OHe+WIMPs multicomponent dark matter scenarios
naturally follow from AC model \cite{FKS} and can be realized in
models of Walking technicolor \cite{KK2}.

The presented approach sheds new light on the physical nature of dark matter.
Specific properties of composite dark matter and its constituents are challenging for their experimental search.
OHe interaction with matter is an important aspect of these studies.
In this context positive result of DAMA/NaI and DAMA/LIBRA experiments may be a signature for exciting phenomena of O-helium nuclear physics.

%%%%%%%%%%%%%%%%%%%%%%%%%%%%%%%%%%%%%%%%%%%%%%%%
%% BACKMATTER
%%%%%%%%%%%%%%%%%%%%%%%%%%%%%%%%%%%%%%%%%%%%%%%%

\begin{theacknowledgments}
  We express our gratitude to Jonathan J. Dickau for kind invitation to contribute this issue of Prespacetime Journal focusing on Cosmology and Gravity.
\end{theacknowledgments}



\bibliographystyle{aipproc}   % if natbib is available

\bibliography{sample}

\IfFileExists{\jobname.bbl}{}
 {\typeout{}
  \typeout{******************************************}
  \typeout{** Please run "bibtex \jobname" to optain}
  \typeout{** the bibliography and then re-run LaTeX}
  \typeout{** twice to fix the references!}
  \typeout{******************************************}
  \typeout{}
 }

\end{document}